%% file: privacy-swap.tex
\documentclass{article}
\usepackage{arxiv}

\usepackage[utf8]{inputenc} 
\usepackage[T1]{fontenc}    
\usepackage{hyperref}       
\usepackage{url}            
\usepackage{nicefrac}       
\usepackage{microtype}      
\usepackage{lipsum}

\usepackage{amsmath,amsfonts,amssymb}
\usepackage{graphicx}
\usepackage{paralist}
\usepackage{color}
\usepackage{balance}
\usepackage{mathtools}
\usepackage{balance}
\usepackage{tabularx}
\usepackage{listings}
\usepackage{hhline}
\usepackage[linesnumbered, boxed]{algorithm2e}
\usepackage{bm}
\usepackage{cite}

\graphicspath{{fig/}}

\newcommand{\para}[1]{\smallskip\noindent{\textbf{#1}}}

\begin{document}

\title{Private and Atomic Exchange of Assets over Zero Knowledge Based Payment Ledger}

\author{
  Zhimin Gao \\
  Department of Computer Science\\
  Auburn University at Montgomery, AL, USA \\
  \texttt{mtion@msn.com} \\
   \And
  Lei Xu  \\
  Department of Computer Science \\
  University of Texas Rio Grande Valley, TX, USA \\
  \texttt{xuleimath@gmail.com} \\
  \And
  Keshav Kasichainula \\
  Department of Computer Science \\
  University of Houston, TX, USA \\
  \texttt{kasikeshav@gmail.com} \\
  \And
  Lin Chen \\
  Department of Computer Science \\
  Texas Tech University, TX, USA \\
  \texttt{chenlin198662@gmail.com} \\
  \And
  Bogdan Carbunar \\
  Department of Computer Science \\
  Florida International University, FL, USA \\
  \texttt{carbunar@gmail.com} \\
  \And
  Weidong Shi \\
  Department of Computer Science \\
  University of Houston, TX, USA \\
  \texttt{larryshi@ymail.com} \\
}

\maketitle
\input{sec-abstract}
\maketitle


\input{sec-introduction}


\input{sec-related-work}

\input{sec-overview}

\input{sec-definition}

\input{sec-protocol}


\input{sec-detail-design}

\input{sec-security-proof}

\input{sec-conclusion}

\bibliographystyle{unsrt}  
\bibliography{ref}

\end{document}

%% file: sec-abstract.tex
\begin{abstract}
{Bitcoin brings a new type of digital currency that does not rely on a central system to maintain transactions. By benefiting from the concept of decentralized ledger, users who do not know or trust each other can still conduct transactions in a peer-to-peer manner. Inspired by Bitcoin, other cryptocurrencies were invented in recent years such as Ethereum, Dash, Zcash, Monero, Grin, etc. Some of these focus on enhancing privacy for instance crypto note or systems that apply the similar concept of encrypted notes used for transactions to enhance privacy (e.g., Zcash, Monero). However, there are few mechanisms to support the exchange of privacy-enhanced notes or assets on the chain, and at the same time preserving the privacy of the exchange operations.  Existing approaches for fair exchanges of assets with privacy mostly rely on off-chain/side-chain, escrow or centralized services.  Thus, we propose a solution that supports oblivious and privacy-protected fair exchange of crypto notes or privacy enhanced crypto assets. The technology is demonstrated by extending zero-knowledge based crypto notes. To address ``privacy'' and ``multi-currency'', we build a new zero-knowledge proving system and extend note format with new property to represent various types of tokenized assets or cryptocurrencies. By extending the payment protocol, exchange operations are realized through privacy enhanced transactions (e.g., shielded transactions). Based on the possible scenarios during the exchange operation, we add new constraints and conditions to the zero-knowledge proving system used for validating transactions publicly.} 
\end{abstract}

\keywords{blockchain, swap, fair exchange, privacy}

%% file: sec-introduction.tex
\section{Introduction}

Cryptocurrency systems based on decentralized ledgers, require each participant to keep a local copy of the transaction history, in order to prevent double-spending without relying on a centralized party. This design however exposes users to privacy concerns, as all transaction information becomes publicly available ~\cite{DBLP:journals/iacr/RonS12}.


To address this problem, several solutions (e.g., Cryptonote~\cite{van2013cryptonote}, Zcash~\cite{sasson2014zerocash}, Monero~\cite{noether2015ring}, and Grin~\cite{henry2018blockchain}),
proposed the use of encrypted notes (or UTXOs), and
focus on privacy issues at the individual payment transaction level, by preventing the disclosure of payment related information, e.g., the recipient wallet address, and the amount of cryptocurrency transferred.
However, privacy preserving or oblivious fair exchange of crypto coins/assets, which is essential for a thriving ecosystem, have been so far largely ignored.

In this paper, we develop a unified framework to support both privacy enhanced payment transactions and a fair exchange of crypto assets without using centralized mixing services, escrow based or off-chain/site-chain approaches~\cite{khalilov2018survey}.
By exchange we mean giving a certain digital asset and receiving another asset in return. The giving and the receiving can be of identical types, e.g., Alice exchanges a ten-dollar note in several one-dollar notes with Bob. They can also be different types of digital assets, for instance, the exchange between one type of crypto asset with another type of crypto asset. Such exchanges are essential for ICOs, as investors need to buy a newly created currency using other currencies. In 2018 alone, there were more than 1,000 ICOs, with a total value of exchange transactions of about 8 billion USD~\footnote{\url{https://www.icodata.io/stats/2018}.}

More specifically, we propose a ledger-based multi-asset system, designed to support both the private exchange (swap) of crypto-assets and regular spending operations. Our solution aims at satisfying several requirements. First, the exchange should to be atomic, i.e., either both or none of the transfers are executed. Second, the system should support exchange between different types of assets, which we call {\it private notes}.
Third, the exchange needs to be private, i.e., information posted on the blockchain should not leak the identities of the parties involved in the exchange, or the nature of the exchange (e.g., types and amount of assets, status of the success of the exchange).
The proposed scheme introduces the concept of sibling note and negative value to allow the linkage of discrete steps in an exchange to guarantee the fairness, and leverages zero-knowledge proof to hide information of exchange transactions while allowing the public to verify the validity. Since a different zero-knowledge proof is required for each step in the exchange, we also utilize a mechanism to merge different types of zero-knowledge proofs into a single format so an adversary cannot learn even the type information of a step in the exchange process.


In summary, the paper provides the following contributions:

\begin{itemize}
    \item We develop a unified and formal framework of privacy preserving exchanges and payment transactions of crypto assets on a decentralized ledger that satisfies security and privacy requirements;
    \item We propose a concrete construction of privacy-preserving asset exchange scheme that supports multiple types of crypto assets and meets the formal definitions; and
    \item We describe and discuss implementation details of the designed scheme using zero-knowledge based cryptocurrency as an exemplary platform.
\end{itemize}




%% file: sec-related-work.tex
\section{Related Work}\label{sec-related-works}

In this section, we briefly review related works. Typically, the exchange of privacy coins can be completed using a centralized intermediary or escrow service provider. The escrow services are basically third party implementations which allow for fast off-chain completion of the transaction and it has been also implemented in TumbleBit~\cite{pc-2}. 



\para{Information exchange}.
For fair information exchange, Alice and Bob exchange their secret information in a fair way: either they learn each others secret at the same time, or they both keep their secrets undisclosed. It has been proved that this is not feasible without the involvement of a third party~\cite{pagnia1999impossibility}, and some works have been done on utilizing the blockchain as a broker to facilitate such exchange~\cite{dziembowski2018fairswap}. An important difference between information exchange and crypto-asset exchange is that the latter requires the transfer of ownership, which is more than fair information exchange. 
For example,
\begin{itemize}
    \item If Alice and Bob exchange information of their assets, they still own their own assets and an extra protocol is required to enforce the transferring of ownership;
    \item If Alice and Bob exchange information that controls the assets (e.g., private key to spend the asset), they are facing a racing condition as both of them still hold their own information after the exchange.
\end{itemize}
Furthermore, most information exchange methods do not consider the requirement of privacy. Without loss of generality, in this paper, crypto assets include cryptocurrencies and tokenized assets. Exchange can occur between any type of digital assets secured by a distributed ledger based transaction system. Further, we assume an environment of privacy assets, for instance, systems based on encrypted notes. The notes and payment transactions are protected with privacy enhancement such as zero-knowledge proving protocols. Towards these objectives, we extend data models and protocols of existing privacy coin schemes, for instance, Zcash/Zerocoin by introducing asset types as an extension to note format for supporting multiple asset types. In addition, we modify the data models and transaction protocols to support oblivious and fair exchanges of crypto notes between users. Our goal is to enable the exchange of privacy enhanced crypto assets with the following main characteristics:  exchange operations are atomic; transactions are publicly verifiable; they are oblivious without disclosing information of the exchange operations.

\para{Zero knowledge execution ($Z_\textit{EXE}$).} 
Zero knowledge can protect the privacy from not only the transaction information but also computation (if it exists). Sean Bowe \emph{et al.} propose a decentralized-ledger-based system that allows transactions to hide all information about the offline computations but any node can still validate the computations even without whole details~\cite{bowe2018zexe}. A use case of cryptocurrency exchange is mentioned in their research. It leverages $Z_{EXE}$ and supports exchange. However, the system needs to execute a script to verify a preset list of conditions before sending the notes. In another word, it has to wait until it gets expected results from the commitment pool. In contrast, in our approach, notes are sent right away. It reduces the delay of transactions. And we have seen a lot of implementations of the Zero-Knowledge proof in different areas like in the Privacy preserving Auditing\cite{pc-13}. There also have been research on having a confidential asserts, scheme where a single node can handle multiple assert types\cite{confidentail_assert}

\para{Decentralized exchange.} Zcash\cite{sasson2014zerocash} and Zcoin\cite{miers2013zerocoin} represent a few of the implementations of the zero-Knowledge proof for privacy where connection between the sender and receivers is cut off. It leaves no trails of both the entities involved in payment transaction\cite{sasson2014zerocash}. In recent times, we have seen several exchange systems built over the blockchain that somewhere in their systems implements the zero-knowledge proof. TEX \cite{tex2019dblp} is one of the proposed exchange systems which claims to be better than the existing decentralized exchanges (DEX).  The proposed TEX system operates in two layers and proposes to execute a trade without the requirement of withholding assets. Other than the exchange mentioned above we have seen a rising number of exchange systems being introduced, and one of them is the Zerocoin Decentralized Exchange (ZDex) which is built over the PIVX platform which in turn, in theory, is a cold fork of the Dash. In its current state, it can support the Bitcoin cash, Litecoin, Dash, Zcash. However, one of the significant drawbacks in this implementation that we have gone ahead and addressed in our approach is that ZDex in its current state doesn't support atomic swap of digital assets.
There are also works on utilizing trusted computing technology to improve the performance and security of smart contract~\cite{das2019fastkitten}, which can be leveraged to implement exchange or even provide privacy protection. However, this type of approaches rely on the trust of the hardware and the vendor behind.

%% file: sec-overview.tex
\section{Overview of Publicly Verifiable Oblivious and Fair Exchange}\label{sec-overview}

In this section, we provide an overview of the proposed privacy preserving exchange mechanism on a decentralized ledger. 

\subsection{Problem Statement}

At a high level, the system is aimed to meet the following requirements. 

\begin{itemize}
\item Decentralized privacy ledger with multi-asset support (for instance zero-knowledge based system with the capability to support multiple types of coins, currencies, or tokens). 

\item Support of privacy protected payment transactions similar to Zcash/Zerocoin (e.g., shielded to shielded payment transactions). 

\item Publicly verifiable means that the public should be able to verify the validity of all transactions including privacy protected payment transactions. 

\item Fair exchange.  Fairness means that an asset exchange operation between two users is atomic. Although the operation may involve multiple payment steps. At the end, either the exchange succeeds with ownership swap of each participant$’$s digital asset, or the operation terminates so that each participant still keeps his/her own asset before the exchange. 

\item Oblivious exchange. Oblivious means that all transactions (both payment transactions and exchange transactions) recorded on the ledger and verified by the public should appear to be identical (indistinguishable from one another) and reveal no information or knowledge about the exchange operation including: the presence of exchange operation, users/accounts involved in the exchange, types of assets and corresponding amounts in the exchange. This means that exchange transactions should be hidden from the shielded payment transactions. Based on history of transactions, an adversary should not be able to separate transactions for fulling exchanges from other shielded payment transactions.     

\item Unified transaction system for both payment and exchange operation. The system should not use separate protocols or algorithms that are unique for payment transactions and exchange operations. All transactions should be verified using the same mechanism and an identical procedure.   

\end{itemize}

We assume that there are more than one type of crypto assets in the system and owners of two different types of assets, Alice and Bob, who do not fully trust each other decide to exchange the ownership of digital assets (e.g., privacy coins, privacy tokens) on the decentralized ledger. 

If privacy is not a concern, we can easily implement a fair exchange operation using a smart contract, and the distributed ledger can guarantee the atomic property of the operation, i.e., either the operation succeeds that Alice and Bob get the other's digital asset, or the operation fails that each one still keeps his/her own assets. However, such an approach does not provide any privacy protection. Every node in the system can see the exchange information. 

It is worth mentioning that all transactions involved in an exchange operation should be publicly verifiable (verified on the chain by nodes of the ledger network using consensus) and all transactions should be oblivious. In other words, all transactions, regardless of their nature (e.g., payment transactions, different steps of exchange operations) should be indistinguishable to the public who verify them. These requirements exclude alternatives such as setting an off-chain payment channel for exchange.  

To simplify discussions and experiments, we use zero-knowledge proof based privacy coin (e.g., Zcash, Zerocoin) as baseline design. We extend the privacy coin data models and proving systems to support multiple assets, and oblivious fair exchange. Exchange is realized by a pre-defined sequence of privacy protected (shielded) payment transactions.   


\begin{figure*}
    \centering
    \includegraphics[width=5.5in]{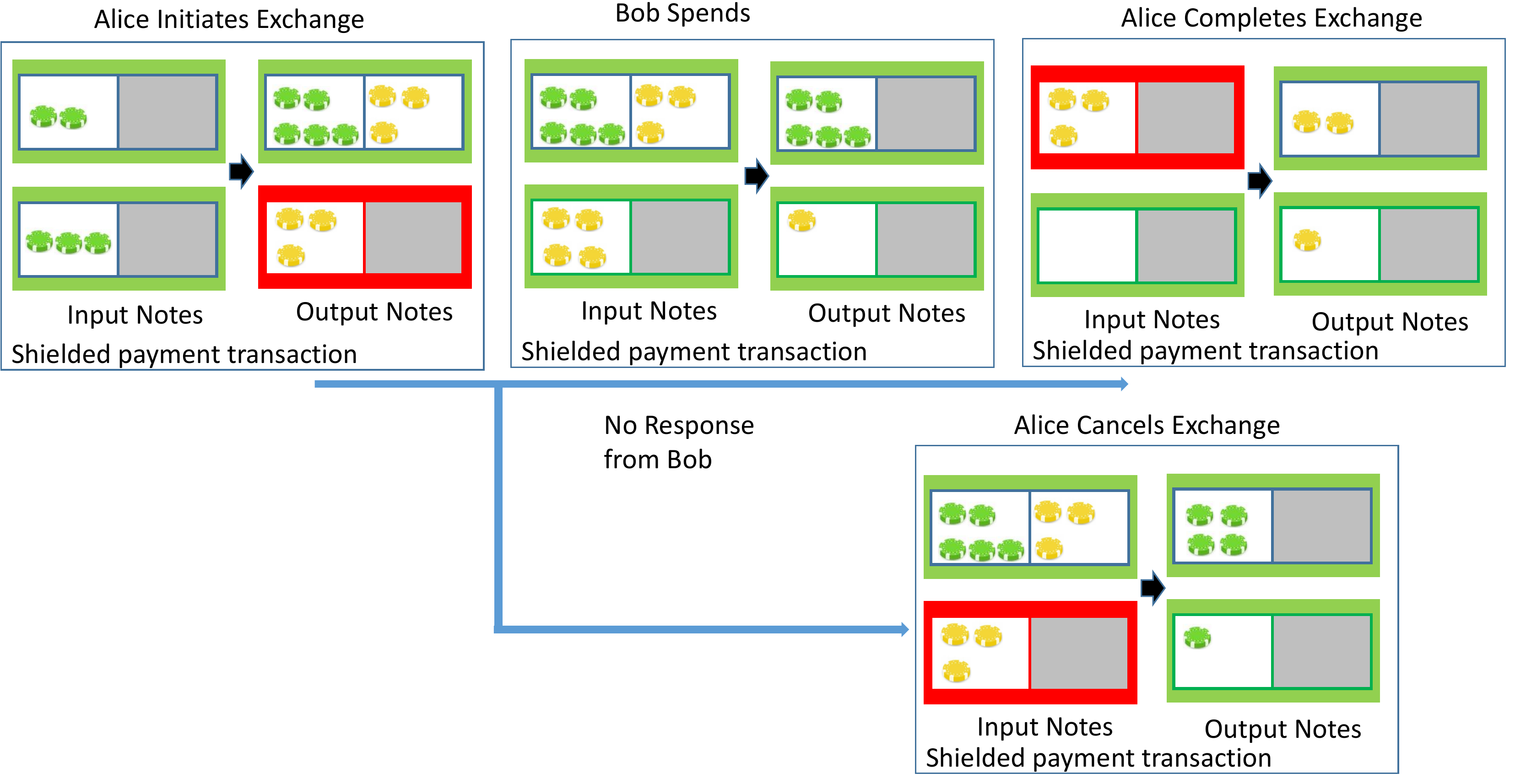}
    \caption{Example transaction sequence between Alice and Bob to complete exchange of assets.}   
    \label{fig-example-swap}
\end{figure*}

\subsection{Zero Knowledge Proof based Currencies}


A zero-knowledge proof of knowledge is a protocol in which a prover can convince a verifier that some statement holds without revealing any information. Roughly speaking, a zero-knowledge proof involves two parties, the {\it prover} and the {\it verifier}. For a statement, the {\it prover} can generate a proof to convince the {\it verifier} of the correctness of the statement. In this process, the {\it verifier} cannot learn anything except the fact that the statement is true (zero-knowledge feature). Zero-knowledge proofs can be either interactive or non-interactive. Interactive zero-knowledge proofs ~\cite{goldwasser1985knowledge,goldreich1996construct} requires the {\it prover} to communicate with the {\it verifier} over multiple rounds to finish the proof. Non-interactive zero-knowledge proofs (NIZK)~\cite{blum1988non,groth2006perfect,de2001robust} do not require multiple rounds of interactions between {\it prover} and {\it verifier} and are more suitable for scenarios where it is difficult for these parties to be online at the same time.

An important tool related to NIZK is zero-knowledge succinct non-interactive arguments of knowledge (zk-SNARK)~\cite{gennaro2013quadratic, ben2013snarks, ben2014succinct}. We refer the readers to ~\cite{ben2013snarks} for a formal definition of zk-SNARK. Leveraging practical ZK-SNARK, Zcash implements a Decentralized Payment System with strong protection of transaction integrity and anonymity built from collision-resistant hash (CRH) functions, commitment schemes, and pseudo-random functions. 
When a crypto-note (UTXO) is spent, the spender needs to create a zero-knowledge proof that there is a corresponding note commitment in the pool of commitments without disclosing which one it is. The proof also shows that the transaction is consistent and that the user knows all the secret keys, without revealing any additional information. 

\subsection{An Example of Exchange Operation}

The proposed oblivious and fair exchange scheme extends Zcash data models, for instance, crypto-note definition, referred as a note for simplification, and the payment protocol. This subsection describes the high-level design concept with a concrete exchange example. 

To support multiple assets, the note format is expanded to include a new integer attribute that specifies the types of digital assets (e.g., coins, tokens). Here we use colors to represent different types of assets (e.g., red coin, green coin, yellow coin). In addition, there are two types of notes, called primary note and sibling note. In \figurename~\ref{fig-example-swap}, note with a green surrounding box is a primary note; and note with a red surrounding box is a sibling note. A sibling note is uniquely associated with a primary note. Each payment transaction includes input notes and output notes where input notes are consumed and output notes are created.  A primary note can be consumed by itself in a payment transaction while the associated sibling note can only be consumed either after the primary note has been consumed already in the past or together with the primary note in a transaction. 

Furthermore, each note contains a debt part (the type of asset and amount) which requires that when the note is consumed in a transaction, the debt part has to be canceled out by additional note with sufficient value that matches with the debt (both asset type and amount). For instance, if a note contains a debt of five red coins, when it is consumed (spent), the user has to mix it with a note containing at least five red coins as inputs.  

Assume that Alice wants to exchange five green coins with Bob$'$s three yellow coins. The sequence of actions may be the following:  

\begin{itemize}

\item Alice initiates the exchange operation by converting her assets into a pair of notes. In the first transaction, she provides two input notes (one with three green coins and one with two green coins). The transaction outputs two notes, one primary note with five green coins and three yellow coins as debt; and a sibling note with three yellow coins. Nodes in the ledger network verify the transactions and store the notes on the ledger.

\item The primary note is encrypted and shared between Alice and Bob. Either Alice or Bob can spend the primary note. 

\item If Bob wants to complete the exchange, he issues a new transaction using the received primary note from Alice and one of his notes as inputs. In this case, his input note contains four yellow coins. The transaction produces two output notes, one with five green coins and the second one with one yellow coin (after canceling the three yellow coin debt embedded in the note received from Alice). 

\item After Bob spends the note sent by Alice, Alice can spend the associated sibling note that contains three yellow coins, which completes the exchange. 

\item In case that Bob does not spend the note received from Alice within a bounded time limit, Alice can cancel the exchange by creating a new transaction that spends both the primary note and the associated sibling note, which cancels out the three yellow coin debt. The output notes will contain the five green coins that she puts in the primary note shared with Bob. As a result, the exchange is terminated. Bob can no longer spend the primary note received from Alice. 

\end{itemize} 

Here in this section, we skip discussions on how the protocol can protect the privacy of each transaction step, ensure fairness and exchange atomicity, and prevent cheating by either Alice or Bob. Details of the protocol design and verification algorithms, as well as security analyses, are provided in the following sections of the paper.

%% file: sec-definition.tex
\section{Definition of Oblivious Multi-Asset Protocol (OMAP) with Fair Exchange Support}\label{sec-definition}

The notations of OMAP are based on extending the notations in Zcash ~\cite{miers2013zerocoin,hopwood2016zcash}. In this paper, for simplification, we apply zk-SNARK and original Zcash design ~\cite{miers2013zerocoin,hopwood2016zcash} to achieve exemplary implementation of OMAP. However, it is worthwhile pointing out that it is possible to realize OMAP using other zero-knowledge proof systems, for instance, zk-STARK ~\cite{cryptoeprint:2018:046}, bulletproof~\cite{bunz2018bulletproofs}, or alternative approaches such as ~\cite{cryptoeprint:2018:176}. Discussion applying these different systems is orthogonal to the scope of this work, which focuses primarily on protocol design of privacy preserving multi-asset transaction and exchange system.  

\subsection{Data Models}

We extend the definition and data format of Zerocoin and Zcash. The notations are based on Zcash specifications. CRH stands for collision-resistant hash algorithm and PRF stands for pseudo-random function.  

\begin{figure}
    \centering
    \includegraphics[width=4.5in]{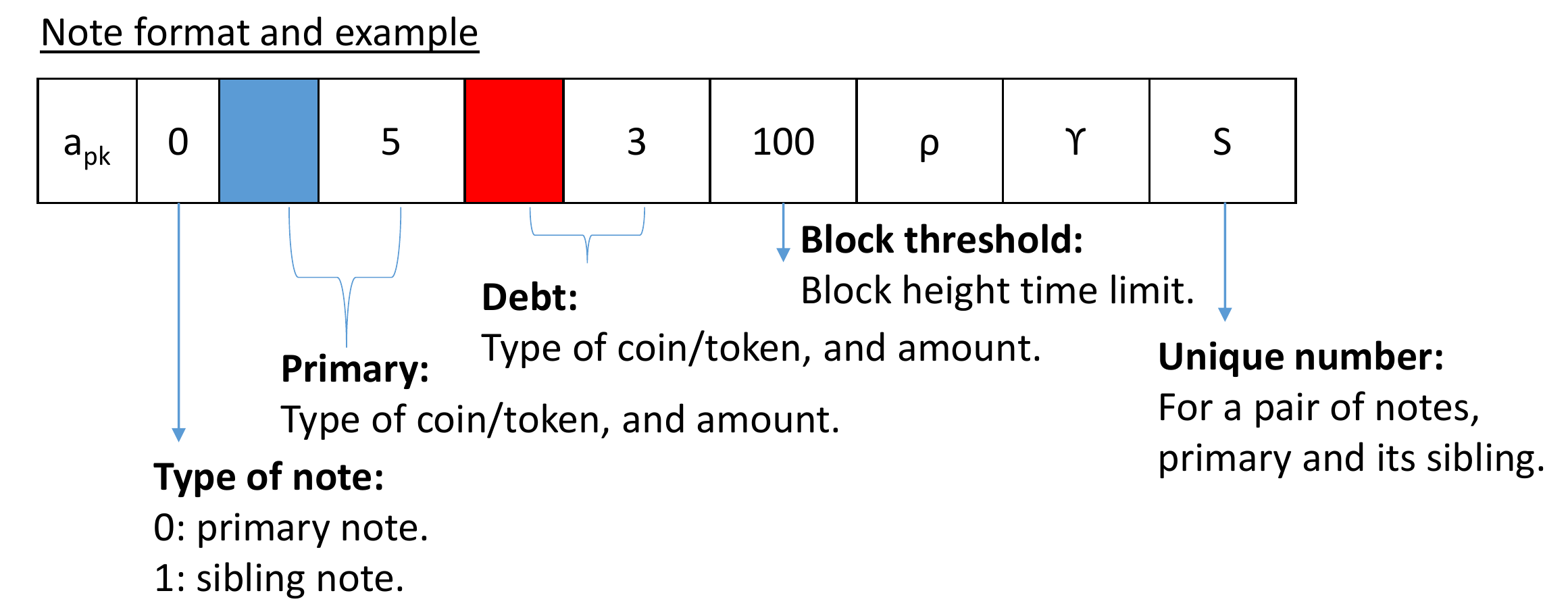}
    \caption{Private note structure.}
    \label{fig-note-structure}
\end{figure}

\para{Multi asset ledger:} There is a multi-asset ledger, $L_{OMAP}$, which records a sequence of transactions in append-only mode. The ledger supports multiple types of tokens or assets using extended note format, described below. The ledger comprises ordered blocks created from a genesis block under a consensus mechanism. Each block has a block height. Further, we assume that blocks are generated with a relatively constant speed. Details of consensus mechanism and block generation can be either based on or follow Zcash design. Although our experiments are based on Zcash implementation, OMAP and $L_{OMAP}$ can be adapted to any distributed ledger based systems such as one based on PoS.  

\para{Public parameters:} In the case of the experiments in this paper, the system uses the same set of public parameters $pp$ as Zcash design. They are generated either by a trusted party at the beginning or through a Multi-Party Computation ceremony ~\cite{cryptoeprint:2017:602,cryptoeprint:2017:1050}. If OMAP is implemented over a proving system that doesn't require trusted setup, the public parameters can be based on common public strings appropriate for the proving system. 

\para{Payment address and note format:} In this work, we use the same design of payment address pair ($a_{sk}$, $a_{pk}$) where $a_{sk}$ is used as spending key. For receiving shielded payment, a user needs to scan ledger $L_{OMAP}$ using  $a_{sk}$. The algorithm is similar to the one described in Zcash blockchain scanning. For each type of asset or token, there is a $v_{pub}$. In this experimental implementation, the value is stored in levelDB. 

An OMAP note, $n$, is similar to Zcash note with extensions to support multiple assets and oblivious exchange. A note is associated with the following attributes:

\begin{itemize}
\item $a_{pk}$: spending key
\item $s$: type of note (primary or sibling note)
\item $color_{1}$: color of primary coin (type of asset)
\item $v_{1}$:  value of primary coin
\item $color_{2}$: color of debt coin (type of asset)
\item $v_{2}$:  debt coin amount
\item $bt$: block height threshold
\item $\rho$: used to compute nullifier (disclosed to the public after spending) 
\item $\gamma$: trapdoor 
\item $S$: unique value for a pair of matching notes
\item $cm$: note commitment
\end{itemize}

\figurename~\ref{fig-note-structure} shows an example note. Attributes including 
$a_{pk}$, $\rho$, $\gamma$, $cm$ are the same as defined by Zcash design. For each note, attribute $s$ specifies the type of note (primary or sibling). For simplifying the discussion, we refer the type of asset as color, for instance, black coin, red coin, which means that black coin and red coin represent different types of assets. The type and amount of primary asset are represented as:  $color_{1}$ and $v_{1}$. Size of $color$ field determines the maximum number of assets supported by OMAP. For each note, there is a second pair of asset type and the amount that represent debt embedded in a note, ($color_{2}$, $v_{2}$). When both $color_{2}$ and $v_{2}$ are zero, it means that the note is a regular note. When both $color_{2}$ and $v_{2}$ are not zero, it means that when the note is spent, it has to be paired with a note with primary asset type matching with $color_{2}$ and the amount greater than or equal with $v_{2}$. The second pair ($color_{2}$, $v_{2}$) is introduced for the purpose to support the atomic exchange of different types of assets between two OMAP users. 

\begin{figure}
    \centering
    \includegraphics[width=4.0in]{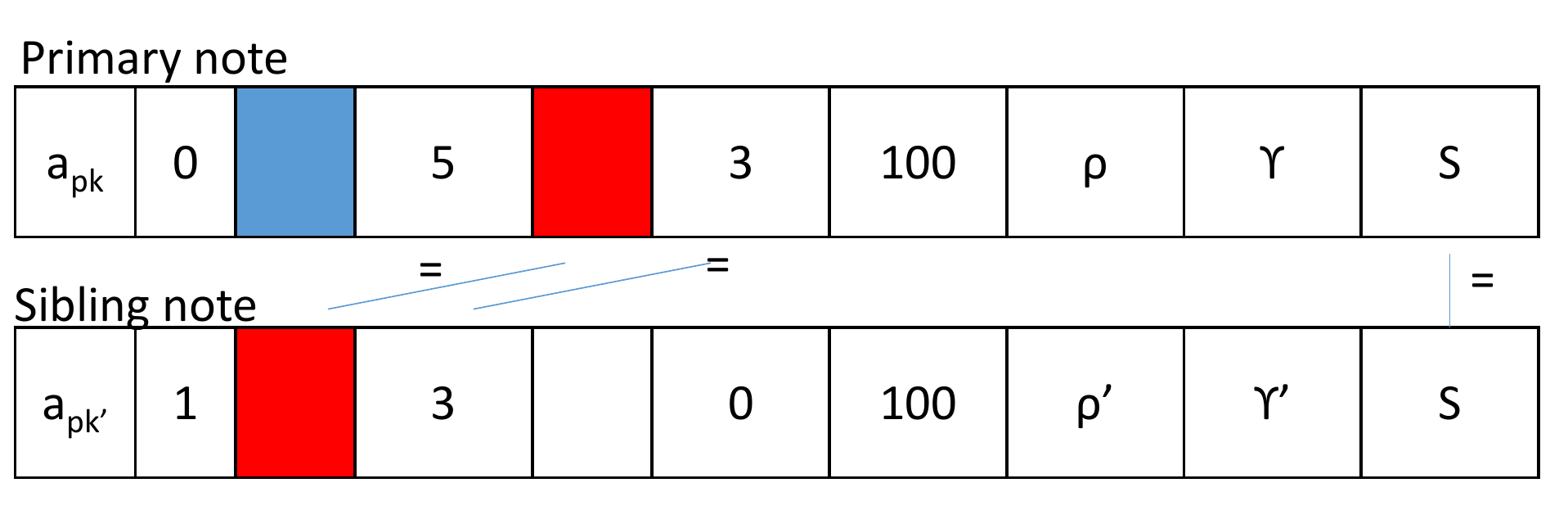}
    \caption{A pair of primary and sibling notes.}   
    \label{fig-note-pair}
\end{figure}

Attribute $bt$ is a block height threshold, which is used for setting a time limit of asset exchange or swap. It is a threshold value that can be configured to decide when the counterparty in exchange has to spend a received note in order to move forward to the next step. When a user initiates an exchange, the user will create a note pair, see example in \figurename~\ref{fig-note-pair}.  One note has $s$ set as zero (primary note) and the second one with $s$ set to 1 (sibling note).  Attribute $S$ is a unique value for each pair of matching notes, which means that a primary note and its sibling note share the same value $S$. Calculation of $S$ is based on a CHR with a unique random input computed from the input notes (specific to a transaction). In addition, for each pair of notes (primary and its sibling), the value of $bt$ needs to be the same; 
$color_{1,2}$ = $color_{2,1}$ $\wedge$ $v_{1,2}$ = $v_{2,1}$; and $v{2,2}$ = 0.

When a note is spent, a nullifier value, $nf$, will be created using attribute $\rho$ as input where $nf$ is determined by $PRF^{nf}_{a_{sk}}(\rho)$. OMAP enforces that nullifiers must be unique in order to prevent double-spending. In addition, uniqueness of nullifiers are used to support the atomic exchange of assets (see section \ref{sec-protocol}). 

\para{Note commitment and nullifier Tree}. OMAP uses incremental Merkle tree of fixed depth for note commitments and nullifiers. Different from Zcash where nullifiers are kept only for preventing double-spending, OMAP maintains a joint Merkle tree, $M$, that includes both note commitments and nullifiers. Alternatively, there could be separate note commitment tree and nullifier tree so that there will be two roots, one for note commitment tree and the second one for nullifier tree. In this work, we assume that the two trees are combined.   

Each commit or nullifier has a (path, pos) where pos is position in the tree. Leaf node $M^{MerkleDepth}_i$ is in the tree with given root rt = $M^0_0$, where M $\in$ $(cm, nf)$.

\begin{figure}
    \centering
    \includegraphics[width=4.5in]{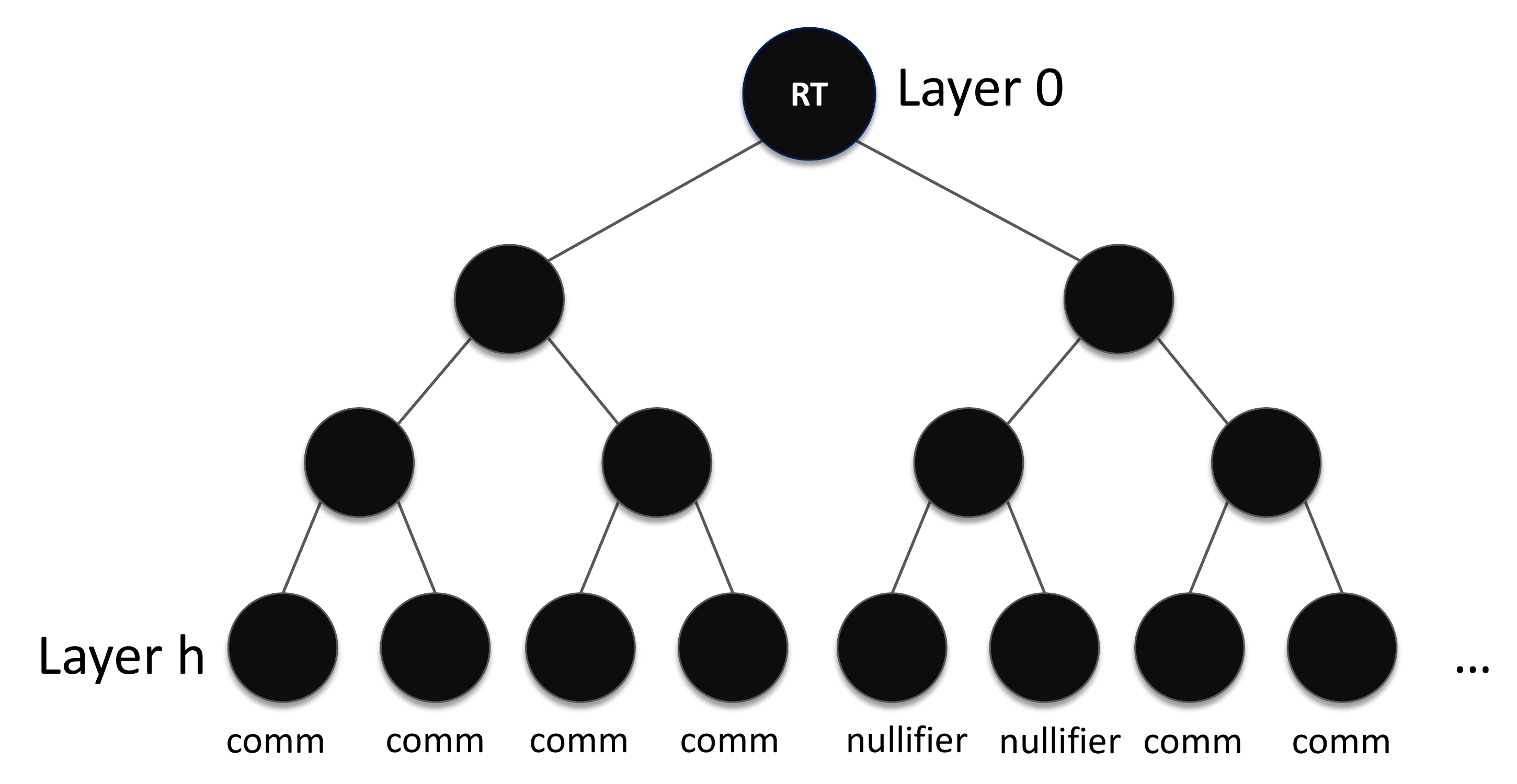}
    \caption{Combined Merkle tree for storing note commitments and nullifiers. Zero knowledge proof is used to show that for any leaf node, there is a path to the root.}   
    \label{fig-tree}
\end{figure}

\subsection{Algorithms}
An Oblivious Multi-Asset Protocol with Exchange Support consists of algorithms ($\textit{Setup}_\textit{OMAP}$, $\textit{CreateAddr}_\textit{OMAP}$,  $\textit{Mint}_\textit{OMAP}$, $\textit{Joinsplit}_\textit{OMAP}$, $\textit{Verify}_\textit{OMAP}$,  $\textit{Receive}_\textit{OMAP}$) defined below. 
 
\para{$\textit{Setup}_\textit{OMAP}(1^\lambda)$:} Based on security parameter $\lambda$,  it creates public parameters $pp_{OMAP}$ defined as  ($pk_\textit{JoinSplit}$, $vk_\textit{JoinSplit}$, $pp_\textit{enc}$, $pp_\textit{sig}$) where $pk_\textit{JoinSplit}$ and $vk_\textit{JoinSplit}$ are a pair of proving and verifying key for JoinSplit transactions (also see section ~\ref{sec-protocol}).

\para{$\textit{CreateAddr}_\textit{OMAP}(pp_\textit{OMAP})$:} Given public parameters $pp_\textit{OMAP}$, it creates payment address pair ($a_{sk}$, $a_{pk}$) where $a_{sk}$ is used as spending key. A user can create arbitrary number of such address pairs.

\para{$\textit{Mint}_\textit{OMAP}(pp_\textit{OMAP}, n, color, v, \pi, L_\textit{OMAP})$:} Given public parameters $pp_\textit{OMAP}$, a note $n$, asset type $\textit{color}$, value $v$, a proof $\pi$, it appends note $n$ to the ledger $L_{OMAP}$, or output $\bot$. 

\para{$\textit{JoinSplit}_\textit{OMAP}(pp_\textit{OMAP}, n^\textit{old}_1, n^\textit{old}_2, a^\textit{old}_{sk,1}, a^\textit{old}_{sk,2}, \textit{color}^\textit{old}_\textit{pub},$ $v^\textit{old}_\textit{pub}, L_\textit{OMAP})$:} It takes as inputs, public parameters $pp_\textit{OMAP}$, Merkle tree root $rt$, two input notes $n^\textit{old}_1$ and $n^\textit{old}_2$, the corresponding spending keys $a^\textit{old}_{sk,1}$ and $a^\textit{old}_{sk,2}$, two output addresses $a^\textit{new}_{pk,1}$ and $a^{new}_{pk,2}$, public value and asset type (color), $\textit{color}^{old}_{pub}$ and $v^{old}_{pub}$, ledger $L_\textit{OMAP}$, it creates two output notes, $n^{new}_1$ and $n^{new}_2$, and a transaction $tr_\textit{JoinSplit}$. 

\para{$\textit{Verify}_\textit{OMAP}$($pp_\textit{OMAP}$, $n^\textit{new}_1, n^\textit{new}_2$, $tr_\textit{JoinSplit}$, $L_\textit{OMAP}$):} Given public parameters $pp_\textit{OMAP}$, Merkle tree root $rt$, a JoinSplit transaction $tr_\textit{JoinSplit}$, two notes $n^\textit{new}_1$ and $n^\textit{new}_2$, and ledger $L_\textit{OMAP}$, it appends $tr_\textit{JoinSplit}$, $n^{new}_1$ and $n^{new}_2$ to the ledger, or output $\bot$.   

\para{$\textit{Receive}_\textit{OMAP}$($pp_\textit{OMAP}$, $rt$, $L_\textit{OMAP}$):} Given public parameters $pp_\textit{OMAP}$, Merkle tree root $rt$, recipient key pair ($a_{sk}$, $a_{pk}$), and the ledger $L_\textit{OMAP}$, it outputs received note $n^\textit{new}$, or output $\bot$.

Note that in this work, we restrict to cases of two input notes and two output notes. Spending cases of more than two input notes or two output notes can be reduced to transactions with two notes as input and output, subject of future extension and research. Similar to the Zcash design, input note or output note can be dummy note (without associated note commitment). In the case of dummy note, the asset type is zero.

$Remark:$ Refer to ~\cite{hopwood2016zcash} for a complete description of notations in the zero-knowledge proving system used by Zcash. The description is based on extending Zcash note format and operations. Here we focus on algorithms and definitions that are extended or modified to support oblivious and fair exchange through $tr_{JoinSplit}$ transactions. Definitions that are not changed can be found in the original Zcash publication and implementation specification.


%% file: sec-protocol.tex
\section{OMAP Protocol}\label{sec-protocol}

\subsection{Preliminary}

The protocol is built by extending the design of zero-knowledge based privacy payment system such as Zcash. These privacy oriented cryptocurrencies apply zero-knowledge proving system as the underlying building block to protect privacy. 
A zero-knowledge proving system is a cryptography protocol that allows proving a particular claim/statement, dependent on two input datasets, public and witness, without disclosing information about the witness input other than that included in the claim/statement, 

A zero-knowledge instance ZK defines ~\cite{hopwood2016zcash}:
\begin{itemize}
\item ZK proving key, ZK.ProvingKey - $pk$
\item ZK verifying key, ZK.VerifyingKey - $vk$
\item a key generation algorithm for creating $pk$ and $vk$  
\item public input, ZK.PublicInput - $\chi$ 
\item witness, ZK.Witness - $\omega$
\item proofs ZK.Proof - $\pi$
\item set of satisfying inputs where ZK.SatisfyingInputs $\in$ ZK.PublicInput $\times$ ZK.Witness 
\item a proving algorithm ZK.Prove that produces ZK.Proof from $pk$ and ZK.SatisfyingInputs
\item a verifying algorithm ZK.Verify that accepts or rejects a statement based on $vk$ $\times$ ZK.PublicInput $\times$ ZK.Proof
\end{itemize}

A zero-knowledge proving system needs to satisfy the following security requirements: 

\para{Completeness:} for security parameter $\lambda$, a $\mathbb{F}$, arithmetic circuit $\mathcal{C}$, and any ($\chi$, $\omega$) $\in$ $\mathcal{R}$ (satisfying inputs), honest prover can convince the verifier with probability $1 - negl(\lambda)$.

\para{Succinctness:} honestly generated proof $\pi$ has $O_n(1)$ bits and Verify($vk$, $\chi$, $\pi$) runs in the $O_\lambda$($|\chi|$). 

\para{Proof-of-knowledge:} If the verifier accepts a proof output by a bounded prover, then the prover knows witness for the given statement. 

\subsection{Transaction Algorithms}

For completeness, this subsection lists the main algorithms. Additional details and definitions can be found in Zcash specification. The main extensions are related to JoinSplit transaction. In addition to one-way payment (shielded to the shielded transaction of Zcash note), JoinSplit is expanded to support additional spending/transaction cases for achieving atomic and oblivious exchange of different types of assets (notes). As described earlier, the design is extended to support multiple types of assets such as a variety of coins or tokens. 

\para{$Setup_{OMAP}$:} It takes $1^\lambda$ as security parameter and outputs public parameters $pp_{OMAP}$ as following. 

\begin{itemize}
\item compute $\mathcal{C}$$_{JoinSplit}$ at security parameter $\lambda$
\item compute ($pk_{JoinSplit}$, $vk_{JoinSplit}$) = KeyGen($1^\lambda$, $\mathcal{C}$$_{JoinSplit}$)
\item create $pp_{enc}$ = $\mathcal{G}$$_{enc}(1^\lambda)$
\item create $pp_{sig}$ = $\mathcal{G}$$_{sig}(1^\lambda)$
\item output $pp_{OMAP}$ = ($pk_{JoinSplit}$, $vk_{JoinSplit}$, $pp_{enc}$, $pp_{sig}$)
\end{itemize}

$pk_{JoinSplit}$ and $vk_{JoinSplit}$ are a pair of proving and verifying key for JoinSplit transactions. In the design and experiment described in this paper, $Setup_{OMAP}$ follows the same Zcash algorithm for constructing public parameters. 

\para{$CreateAddr_{OMAP}$:} It takes public parameters $pp_{OMAP}$ as input and creates a pair of transmission key ($a_{pk}$, $pk_{enc}$) and receiving key ($a_{pk}$, $sk_{enc}$) where a note sent to a recipient is encrypted using $pk_{enc}$ and it is retrieved by the recipient from the ledger using $sk_{enc}$. 

In case of initializing the first JoinSplit transaction for an exchange operation, sender and recipient share the same spending key $a_{sk}$ for spending the primary note. In this case, $a_{sk}$ can be created from a shared secret between the sender who initiates the exchange and recipient, for instance $a_{sk}$ = $PRF^{shared}_{a_{sk}}(shared secret||h_{sig})$ where $h_{sig}$ = $CRH(nf^{old}_1, nf^{old}_2, pk_{sig})$ and PRF can be SHA-256. 

\para{$Mint_{OMAP}$:} It takes public parameters $pp_{OMAP}$, public address pair ($a_{pk}$, $pk_{enc}$), $\pi$,  asset $color$, value $v$ where $color$ $\in$ (1, ..., $color_{max}$) and $v$ $\in$ (0, ..., $v_{max}$). It appends note $n$ to the ledger $L_{OMAP}$, or output $\bot$. This algorithm is used to mint notes of different asset types based on the public values (each asset type and amount) and put the notes on $L_{OMAP}$. 

\para{$JoinSplit_{OMAP}$:} It takes as inputs, public parameters $pp_{OMAP}$, Merkle tree root $rt$, two input notes $n^{old}_1$ and $n^{old}_2$, the corresponding spending keys $a^{old}_{sk,1}$ and $a^{old}_{sk,2}$, two output addresses $a^{new}_{pk,1}$ and $a^{new}_{pk,2}$, public value and asset type (color), $color^{old}_{pub}$ and $v^{old}_{pub}$, ledger $L_{OMAP}$. it creates two output notes, $n^{new}_1$ and $n^{new}_2$, and a transaction $tr_{JoinSplit}$. 

To support asset exchange, there are different spending cases for JoinSplit transaction. Combination of these cases can be applied to achieve atomic exchange of assets between users. The list of JoinSplit spending cases and details are described in the following section. Here we focus on operations and procedures common to all the spending cases. These include:

\begin{itemize}
\item for each input note, compute nullifier $nf^{old}_i$ = $PRF^{nf_{old}}_{a^{old}_{sk,i}}$($\rho^{old}_i$) = SHA-256($a^{old}_{sk,i}\|\rho^{old}_i$) 
\item create signature key pair $(pk_{sig}, sk_{sig})$ = $\kappa_{sig}(pp_{sig})$ where signature scheme is sUF-CMA (e.g., Ed25519)
\item compute $h_{sig}$ = $CRH(nf^{old}_1, nf^{old}_2, pk_{sig})$
\item for each new note, sample $\gamma^{new}_i\overset{R}{\longleftarrow}COMM.Trapdoor$
\item set $\rho^{new}_i$ = $PRF^{\rho}_{\varphi}(i, h_{sig})$ = SHA-256($i\|\varphi\|h_{sig}$)
\item compute note commitment $cm^{new}_i$ = $COMM^{s}_{r^{new}_i}$($a^{new}_{pk,i}$, $s^{new}_i$, $color^{new}_{i,1}$, $v^{new}_{i,1}$, $color^{new}_{i,2}$, $v^{new}_{i,2}$, $bt^{new}_i$, $\rho^{new}_i$, $\gamma^{new}_i$, $h_{sig}$) where COMM can be SHA-256
\item set new note as $n^{new}_i$ = ($a^{new}_{pk,i}$, $s^{new}_i$, $color^{new}_{i,1}$, $v^{new}_{i,1}$, $color^{new}_{i,2}$, $v^{new}_{i,2}$, $bt^{new}_i$, $\rho^{new}_i$, $\gamma^{new}_i$, $h_{sig}$, $cm^{new}_i$)
\item compute old note spending signature $h_{i}$= $PRF^{pk}_{a^{old}_{sk,i}}(i||h_{sig})$
\item encrypt new note as $N^{enc,new}_i$ = $\mathcal{E}_{enc}(pk^{new}_{enc,i}, n^{new}_i)$
\item set public input $\chi$ = ($rt$, $nf^{old}_1$, $nf^{old}_2$, $cm^{new}_1$, $cm^{new}_2$, $v^{old}_{pub,color}$, $v^{new}_{pub,color}$, $block_{n}$, $h_{sig}$, $h_1$, $h_2$)
\item set witness $\omega$ = ($path_1$, $path_2$, $pos_1$, $pos_2$, $n^{old}_1$, $n^{old}_2$, $a^{old}_{sk,1}$, $a^{old}_{sk,2}$, $\varphi$, $dummy^{old}_1$, $dummy^{old}_2$, $n^{new}_1$, $n^{new}_2$, $path3$, $pos3$, $n^{old}_3$, $a^{old}_{sk,3}$, $path4$, $pos4$, $nf^{old}_3$)
\item compute proof $\pi_{JoinSplit}$ = Prove($pk_{JoinSplit}$, $\chi$, $\omega$)
\item set transaction message $m$ = ($\chi$, $\pi_{JoinSplit}$, $info$, $N^{enc, new}_1$, $N^{enc, new}_2$)
\item set message signature $\delta$ = $S_{sig}(sk_{sig}, m)$
\item $tx_{JoinSplit}$ = ($rt$, $nf^{old}_1$, $nf^{old}_2$, $cm^{new}_1$, $cm^{new}_2$, $color^{new}_{pub}$, $v^{new}_{pub}$, $info$, $*$) where
\item * = ($pk_{sig}$, $h_1$, $h_2$, $\pi_{JoinSplit}$, $N^{enc, new}_1$, $N^{enc, new}_2$, $\delta$)
\end{itemize}

In $\omega$, the data fields are defined as: 

\begin{itemize}
\item $path_{1..N^{old}}$ and $pos_{1..N^{old}}$, N=1 or 2: Merkle tree paths of old notes 
\item $n^{old}_{1..N^{old}}$, N=1 or 2: old notes
\item $a^{old}_{sk,1..N^{old}}$, N=1 or 2: old note spending key
\item $\varphi$: random seed
\item $dummy^{old}_{1..N^{old}}$: old note is dummy note or not 
\item $n^{new}_{1..N^{new}}$: new notes
\item $path_3$ and $pos_3$: Merkle tree path and position of associated note with one of the old notes (primary and sibling pair)
\item $n^{old}_3$: associated note with one of the old notes (primary and sibling pair)
\item $a^{old}_{sk,3}$: spending key of the associated note (primary and sibling pair)
\item $path_4$ and $pos_4$: Merkle tree path and position of associated note nullifier (primary and sibling pair)
\item $nf^{old}_3$: associated note nullifier (primary and sibling pair)
\end{itemize} 

When generating $\pi_{JoinSplit}$, the circuit verifies spending case specific constraints that will be described in the next section. There are checks common to all JoinSplit spending cases including: 

\begin{itemize}
\item Merkle path validity: $(path_i, pos_i)$ valid tree path from NoteCommit $(n^{old}_i)$ to rt
\item Nullifier integrity: $nf^{old}_i$ = $PRF^{nf_{old}}_{a^{old}_{sk,i}}(\rho^{old}_i)$ 
\item Spending key validity: $a^{old}_{pk_i}$ = $PRF^{addr}_{a^{old}_{sk,i}}(0)$
\item Uniqueness of $\rho^{new}_i$: $\rho^{new}_i$ = $PRF^{\rho}_{\varphi}(i, h_{sig})$
\item Note commit: $cm^{new}_i$ = NoteCommit$(n^{new}_i)$
\end{itemize}

\para{$Verify_{OMAP}$:} It takes public parameters $pp_{OMAP}$, Merkle tree root $rt$, a JoinSplit transaction $tr_{JoinSplit}$, two notes $n^{new}_1$ and $n^{new}_2$, and ledger $L_{OMAP}$, it appends $tr_{JoinSplit}$, $n^{new}_1$ and $n^{new}_2$ to the ledger, or output $\bot$.  The algorithm does the following: 

\begin{itemize}
\item parse $tr_{JoinSplit}$
\item $b1$ ${\leftarrow}Verify(vk_{JoinSplit}, \chi, \pi_{JoinSplit})$
\item $b2$ ${\leftarrow}\nu_{sig}(pk_{sig}, m, \delta)$
\item output $\bot$ if any of the following is true:
\begin{itemize}
\item $b1$ $\wedge$ $b2$ is false 
\item $nf^{old}_1$ or $nf^{old}_2$ appears on $L_{OMAP}$
\item $nf^{old}_1$ = $nf^{old}_2$ 
\item Merkle root rt not on $L_{OMAP}$
\item $h_{sig}$ doesn't match with $CRH(nf^{old}_1, nf^{old}_2, pk_{sig})$
\end{itemize}
\end{itemize} 

\para{$Receive_{OMAP}$:} Given public parameters $pp_{OMAP}$, Merkle tree root $rt$, recipient key pair ($a_{sk}$, $a_{pk}$), and the ledger $L_{OMAP}$, it outputs received note $n^{new}$, or output $\bot$. The algorithm follows Zcash design. It scans the ledger and outputs received note for each JoinSplit transaction. 

%

%% file: sec-detail-design.tex
\section{Design of Privacy Preserving Fair Exchange}\label{sec-construction}

In this section, we provide the details of a concrete privacy preserving exchange scheme implemented over ZK based privacy multi-asset ledger. A fair exchange between users is achieved through multiple JoinSplit payment transactions of private assets. 

\subsection{Overview of Transaction Workflow}

\begin{figure}
    \centering
    \includegraphics[width=5.2in]{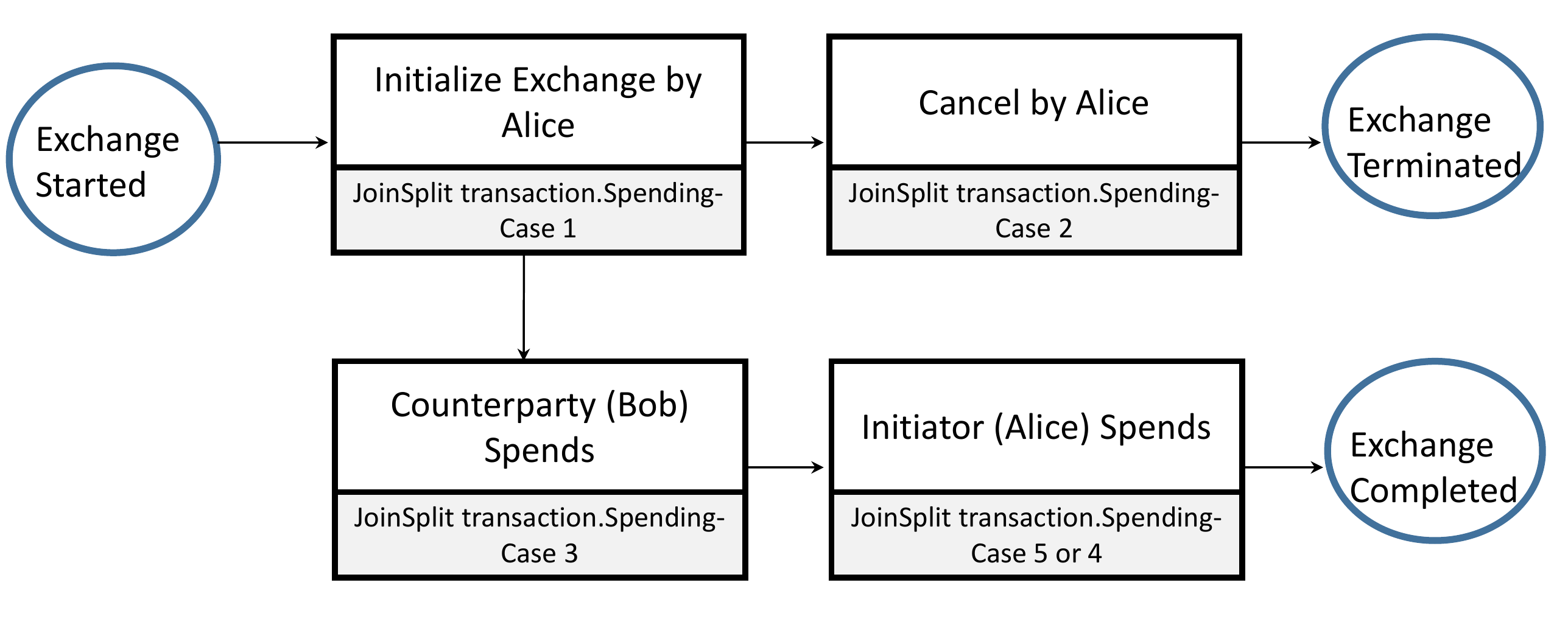}
    \caption{Workflow of privacy-preserving fair exchange of private assets. The fair exchange is achieved through shielded JoinSplit transactions.  Alice initializes the process.  She can cancel the operation (depending on time constraints). After Bob takes action toward completion, Alice can issue a transaction to complete the exchange. If Bob does not respond with a transaction, Alice can cancel the exchange operation and recover her notes.  All transactions are shielded transactions of privacy notes. }
    \label{fig-work-flow}
\end{figure}

Asset exchange is achieved using a sequence of extended JoinSplit transactions. The scheme is designed to ensure fairness, correctness, and privacy of the exchange operations. Fairness means that the entire operation (sequence of JoinSplit payments) satisfies atomicity. Correctness requires that the system should maintain a balance of assets during exchange transactions. Privacy means that all JoinSplit transactions recorded on the ledger and verified by the public should appear to be identical (indistinguishable from one another) and reveal no information about the exchange (e.g., users involved, asset types, amounts, exchange status). 

Atomicity is critical for the privacy-preserving exchange operation. It does not mean that the exchange of assets needs to be completed using one step. It requires the exchange/swap to be fair, i.e., at the end of the exchange, the operation either fails and two parties get their notes back, or succeeds that each party gets the other's note. \figurename~\ref{fig-work-flow} illustrates the workflow of an exchange, where each step involves a shielded JoinSplit transaction. The flow may be affected by actions of the participating parties. Further details of each transaction case are described next. 


\input{subsec-detailed-workflow}

%% file: subsec-detailed-workflow.tex
\subsection{Transactions for Supporting Exchange}\label{subsec-workflow}

OMAP achieves exchange of assets with privacy between two parties using JoinSplit transactions. In order to support fair exchange, the JoinSplit transaction protocol described in Zcash design is extended with new spending scenarios. Accordingly, the zero-knowledge proving system for JoinSplit instance is also extended to include new constraints for each new type of JoinSplit spending scenario. 

OMAP extends note format with new attributes. As a result, it creates multiple JoinSplit spending types. The number of JoinSplit scenarios depends on, type of each input note (primary or sibling), type of each output note (primary or sibling), the value of the debt part in input notes (no debt or with debt). With the restriction of two input notes and two output notes, there are eighteen spending scenarios when ignoring the order of note types. This means that changing the position of a  note within inputs or outputs will not affect the spending scenario (otherwise there will be total of thirty-two spending scenarios. Of the eighteen spending scenarios, there are eight allowed cases. They are shown in Table ~\ref{table-cases}.  The subsequent discussion provides details of each new spending case, in particular, constraints that must be satisfied by the zero-knowledge proof.  

\begin{table}
\centering
\caption{Spending cases based on input and output note types (assume two input notes and two output notes). $\times$ indicates disallowed scenarios. }
\label{table-cases}
\begin{tabularx}{0.9\columnwidth}{|l|l|l|X|} 
\hline
Input Types                       & Debt? & Output Types   & Cases     \\ 
\hline
$<0,0>$                     & No    & $<0,0>$   & Default shielded payment transaction    \\ 
\cline{3-4}
                             &       & $<0,1>$  & Case 1   \\ 
\cline{3-4}
                             &       & $<1,1>$  & $\times$         \\ 
\cline{2-4}
                            & YES    & $<0,0>$  & Case 3    \\ 
\cline{3-4}
                             &       & $<0,1>$  & Case 1           \\ 
\cline{3-4}
                             &       & $<1,1>$ & $\times$          \\ 
\hline
$<0,1>$ & NO    & $<0,0>$ & Case 4                 \\ 
\cline{3-4}
                             &       & $<0,1>$ & Case 4                           \\ 
\cline{3-4}
                             &       & $<1,1>$ & $\times$          \\ 
\cline{2-4}
                             & YES   & $<0,0>$ & Case 5 or Case 2 \\ 
\cline{3-4}
                             &       & $<0,1>$ & Case 2            \\ 
\cline{3-4}
                             &       & $<1,1>$ & $\times$        \\ 
\hline
$<1,1>$ & NO    &                              & $\times$       \\ 
\cline{2-4}
                             & YES   &                              & $\times$                         \\
\hline
\end{tabularx}
\end{table}

\para{Default shielded payment transaction:} This is the case of the original JoinSplit transaction where one user sends notes to another user.  Its definition is identical to JoinSplit transaction described in the Zcash design.  

\para{Exchange initialization:} Without loss of generality, we assume that Alice initializes an exchange process. In this case, the ZK proving algorithm needs to show that the satisfying instance can meet the following constraints in addition to the requirements common to all JoinSplit transactions. Satisfying conditions for the transaction instance include:

$$ \left\{
\begin{aligned}
&s^{old}_1 =s^{old}_2 = 0 \\
&v^{old}_{1,2}  = v^{old}_{2,2}  = 0  || v^{old}_{1,2} >0,v^{old}_{2,2} = 0  || v^{old}_{1,2} =0,v^{old}_{2,2} > 0 \\
&s^{new}_1  = 0, s^{new}_2 = 1 \\
&v^{new}_{1,2}  > 0 \\
&v^{new}_{2,2}  = 0 \\
\end{aligned}
\right.
$$

Satisfying conditions determine specific requirements for the input and output notes during a transaction. During proof generation, it also helps the system to select an appropriate routine (sub arithmetic circuit) to execute. Suppose a note can be denoted simply as $[s, color_1, v_1, color_2, v_2]$; and Alice creates 5 units of green note (color code: 3) to exchange 7 units of red note (color code: 2) with Bob, we have: 

\begin{figure}
    \centering
    \includegraphics[width=4in]{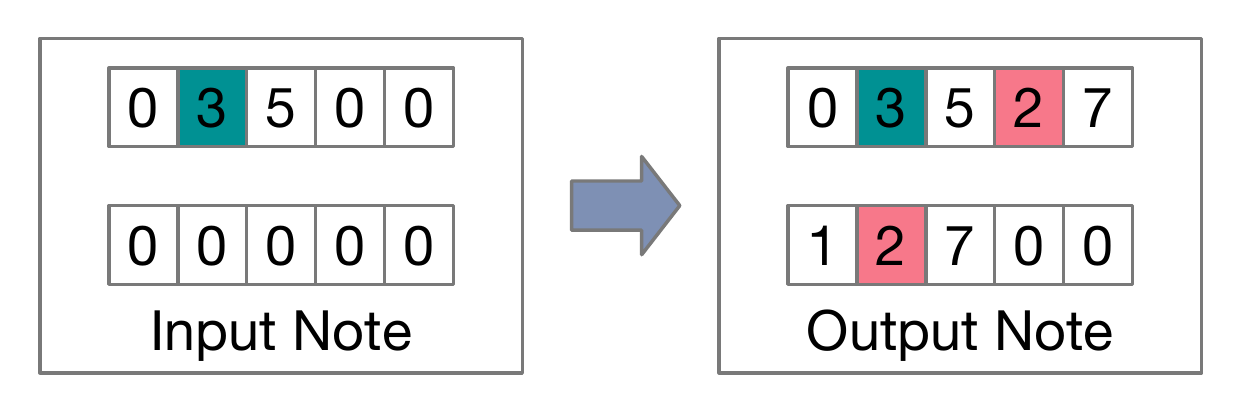}
    \caption{Case 1 transaction example.}
    \label{fig-case-2}
\end{figure}

Note that output note (sibling note) $[1, 2, 7, 0, 0]$ cannot be spent unless output note $[0, 3, 5, 2, 7]$ is spent already or spent at the same time. $[0, 3, 5, 2, 7]$ has a negative part where $color_2=2$ and $v_2=7$. Additional satisfying conditions for this case, besides the original verification of privacy JoinSplit ~\cite{hopwood2016zcash}:
When $v^{old}_{1,2}  = v^{old}_{2,2}  = 0$, and the second note is not a dummy note, the proving algorithm verifies: 

$$ \left\{
\begin{aligned}
&color^{old}_{1,1}=color^{old}_{2,1}=color^{new}_{1,1} \\
&color^{new}_{1,2}=color^{new}_{2,1} \\
&v^{new}_{1,2}=v^{new}_{2,1} \\
&v^{old}_{1,1} + v^{old}_{2,1}=v^{new}_{1,1}
\end{aligned}
\right.
$$

When $v^{old}_{1,2}  > 0$ or $v^{old}_{2,2} > 0$, the proving algorithm verifies similar sets of satisfying conditions as additional sub-cases under this transaction scenario.

Either Alice or Bob can spend the created primary note. This means that both of them have access to the same note spending key. This can be achieved using several approaches. Both can either run a secret sharing protocol offline or encrypt exchanged messages using each other's address public key and store ciphertexts on the ledger or as part of the encrypted memo associated with a note. 

\para{Exchange cancellation by the initiator:} The initiator, Alice, is free to cancel the swap before the other party takes actions (the system can enforce that Alice can only cancel after a pre-determined time limit specified as block height threshold). This is equivalent to the case that Alice starts  a second JoinSplit transaction to get her notes back before Bob does anything. In case of a cancellation transaction, the proving algorithm verifies the following satisfying conditions for the transaction instance:

$$ \left\{
\begin{aligned}
&s^{old}_1  = 0 \\
&s^{old}_2 = 1 \\
&v^{old}_{1,2}  > 0 \\
&v^{old}_{2,2} = 0 \\
&(s^{new}_1=0, s^{new}_2=0, v^{new}_{1,2}=0, v^{new}_{2,2}=0 &or\\
&s^{new}_1=0, s^{new}_2=1, v^{new}_{1,2}>0, v^{new}_{2,2}=0)
\end{aligned}
\right.
$$

If Alice issues a new transaction to recover 5 units of green note and cancel the exchange, we have: 

\begin{figure}
    \centering
    \includegraphics[width=4in]{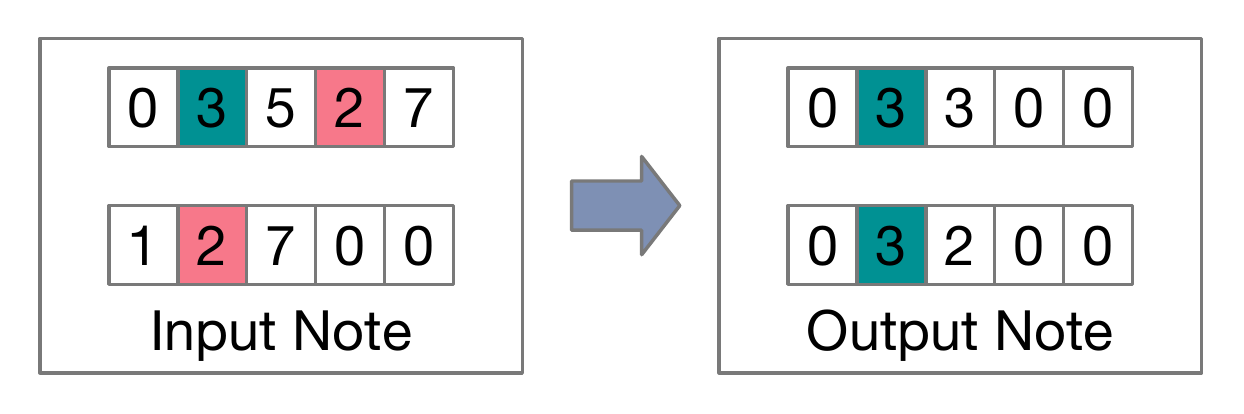}
    \caption{Case 2 transaction example.}
    \label{fig-case-3}
\end{figure}

When $v^{new}_{1,2}  = v^{new}_{2,2}  = 0$, the proving algorithm verifies the sets of satisfying conditions.

$$ \left\{
\begin{aligned}
&color^{old}_{1,2}=color^{old}_{2,1}\\
&color^{new}_{1,1}=color^{new}_{2,1}=color^{old}_{1,1} \\
&v^{new}_{1,2}=v^{old}_{2,1} \\
&v^{new}_{1,1} + v^{new}_{2,1}=v^{old}_{1,1}
\end{aligned}
\right.
$$

Similarly, additional transaction sub-cases cover when $s^{new}_2=1$ and $v^{new}_{1,2}>0$ due to the situation that notes may change order of positions.


\para{Counterparty response with transaction:} In order to spend the note that he receives and complete the exchange, Bob needs to handle the note from Alice with debt. He must spend one of his notes that can cancel out the debt encoded in the note (both type of asset and amount). The requirement is reflected in the following satisfying conditions by the proving algorithm for the transaction instance that Bob creates:

$$ \left\{
\begin{aligned}
&s^{old}_1 = s^{old}_2 = s^{new}_1 = s^{new}_2 = 0 \\
&(v^{old}_{1,2}>0, v^{old}_{2,2}=0 || v^{old}_{1,2}=0, v^{old}_{2,2}>0) \\
&v^{new}_{1,2}=0, v^{new}_{2,2}=0
\end{aligned}
\right.
$$

In this transaction, Bob receives 5 units of green coins by sending 9 units of red coins in order to finish the exchange operation: 

\begin{figure}
    \centering
    \includegraphics[width=4in]{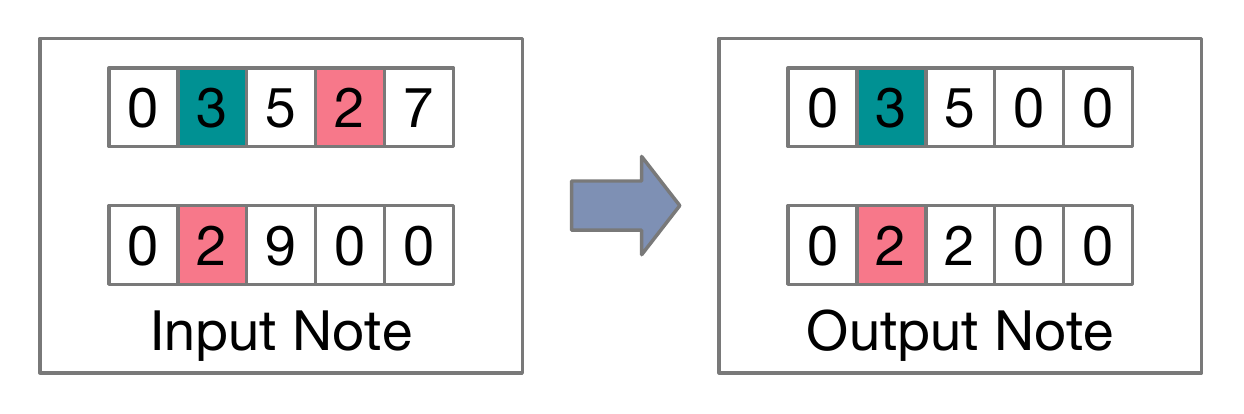}
    \caption{Case 3 transaction example.}
    \label{fig-case-4}
\end{figure}

When $v^{old}_{1,2}>0$ and $color^{old}_{1,1}=color^{new}_{1,1}$, the proving algorithm verifies:

$$ \left\{
\begin{aligned}
&v^{old}_{1,1}=v^{new}_{1,1} \\
&color^{old}_{1,2}=color^{old}_{2,1}\\
&color^{old}_{2,1}=color^{new}_{2,1} \\
&v^{old}_{1,2} + v^{new}_{2,1}=v^{old}_{2,1} \\
&block_n \leq bt^{old}_1
\end{aligned}
\right.
$$

The last condition means that Bob can only spend the note before a block height of $bt^{old}_1$. If not, Alice can cancel the exchange. This prevents the scenario that exchange neither completes nor terminates by a cancellation. The proving algorithm also needs to verify other sub-cases under this spending scenario when the order of notes changes, e.g., when $v^{old}_{1,2}>0$ and $color^{old}_{1,1}=color^{new}_{2,1}$, etc. After Bob spends, a nullifier will be created and inserted to the pool of nullifiers and combined Merkle tree, which Alice can use for completing the exchange. 

\para{Exchange completion by the initiator:} In this scenario, Alice spends after Bob. Alice knows when she can spend by scanning the pool of nullifiers. When she detects the expected nullifier, she can complete the exchange by creating a new transaction to spend the matching sibling note. The proving algorithm verifies the following satisfying constraints by the transaction instance: 

\begin{subequations}
$$ \left\{
\begin{aligned}
&s^{old}_1 = 1 \\
&s^{old}_2 = 0 \\
&v^{old}_{1,2} = v^{old}_{2,2} = 0 \\
&(s^{new}_1 = 0, s^{new}_2 = 0, v^{new}_{1,2}=0, v^{new}_{2,2}=0 & or \\
&s^{new}_1 = 0, s^{new}_2 = 1, v^{new}_{1,2}>0, v^{new}_{2,2}=0) \\
\end{aligned}
\right.
$$
\end{subequations}

Alice completes the exchange with a new transaction:

\begin{figure}
    \centering
    \includegraphics[width=4in]{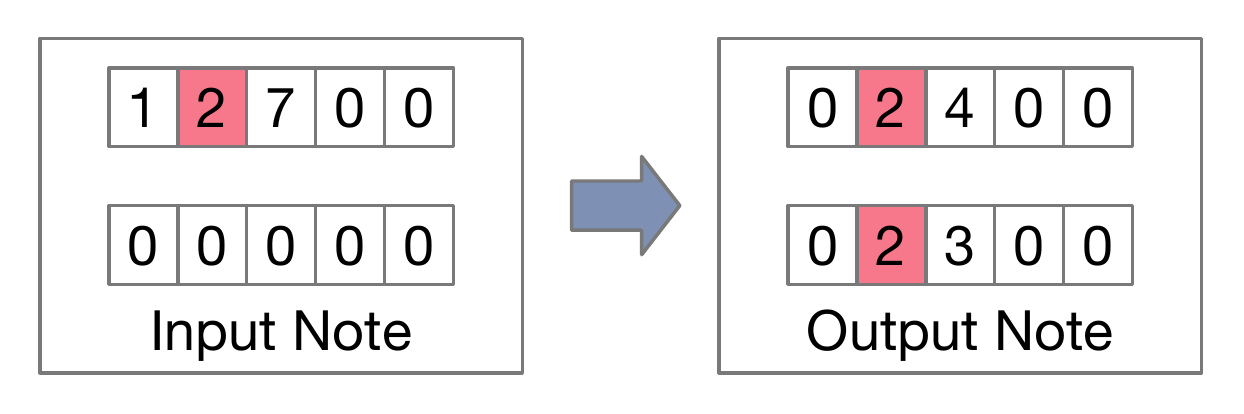}
    \caption{Case 4 transaction example.}
    \label{fig-case-5}
\end{figure}

Note that the first input note can only be spent if there is a matching nullifier for a note already spent by Bob that is the primary note associated with the first input note  (See Case 3 transaction example).
Additional satisfying requirements for this transaction scenario include:

$$ 
\left\{
\begin{aligned}
&color^{old}_{1,1}=color^{old}_{2,1}=color^{new}_{1,1}=color^{new}_{2,1}\\
&v^{old}_{1,1} + v^{old}_{2,1}=v^{new}_{1,1}+v^{new}_{2,1} \\
&block_n > bt^{old}_1\\
&s^{old}_1=1\\
&\exists note n^{old}_3 \wedge (h^{old}_{3,sig}= h^{old}_{1,sig})\\
&(\exists cm^{old}_3\ with\ Merkle\ path\ (path3, pos3)\ respect\ to\ rt\\
& \wedge cm^{old}_3 = NoteCommit(n^{old}_3))\\
&(\exists nf^{old}_3\ with\ Merkle\ path\ (path4,pos4)\ respect\ to\ rt\\
& \wedge nf^{old}_3 = \textit{PRF}^{nf}_{a^{old}_{sk,3}}(\rho^{old}_3))\\
&|pos3 - pos1| =1 \footnotemark 
\end{aligned}
\right.
$$

Similar to Case 3 transaction, one can set a threshold using block height $bt^{old}_1$ to restrict the spending period of the note. In addition, the proving algorithm needs to verify additional sub-cases of transaction instances such as $s^{new}_2=1$ and $v^{new}_{1,2}>0$, etc.

\para{Exchange completion by the initiator (a second scenario):} In this scenario, Alice spends after Bob to complete the exchange. The proving algorithm needs to verify the following constraints on input notes and output notes:
$$ \left\{
\begin{aligned}
&color^{old}_{1,1}=color^{old}_{2,2}=color^{new}_{1,1}\\
&color^{old}_{2,1}=color^{new}_{2,1} \\
&v^{old}_{1,1} = v^{old}_{2,2}+v^{new}_{1,1} \\
&v^{old}_{2,1} = v^{new}_{2,1} \\
&block_n > bt^{old}_1\\
&block_n \leq bt^{old}_2\\
&\exists n^{old}_3 \wedge (h^{old}_{3,sig}= h^{old}_{1,sig})\\
&(\exists cm^{old}_3\ with\ Merkle\ path\ (path3, pos3)\ respect\ to\ rt\\ 
& \wedge cm^{old}_3 = NoteCommit(n^{old}_3))\\
&(\exists nf^{old}_3\ with\ Merkle\ path\ (path4,pos4)\ respect\ to\ rt\\
& \wedge nf^{old}_3 = PRF^{nf}_{a^{old}_{sk,3}}(\rho^{old}_3))\\
&|pos3 - pos1| =1 
\end{aligned}
\right.
$$

Additional conditions that need to be satisfied by this scenario:
When $s^{new}_1=s^{new}_2=v^{new}_{1,2}=0$, the proving algorithm needs to verify:
$$ \left\{
\begin{aligned}
&s^{old}_1 = 1 \\
&s^{old}_2 = 0 \\
&v^{old}_{1,2} =0\\
&v^{old}_{2,2} >0 \\
&s^{new}_1 = 0, s^{new}_2 = 0, v^{new}_{1,2}=0, v^{new}_{2,2}=0\\
\end{aligned}
\right.
$$

\begin{figure}
    \centering
    \includegraphics[width=4in]{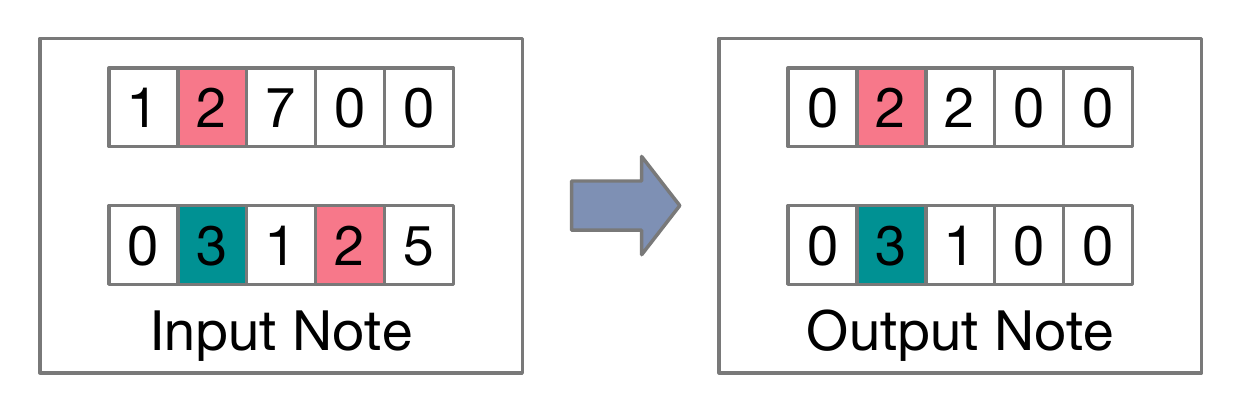}
    \caption{Case 5 transaction example.}
    \label{fig-case-6}
\end{figure}

\footnotetext{\label{note1}The conditions guarantee that note commitment $cm^{old}_3$ is the neighbor of its sibling/pairing note$'$ commitment $cm^{old}_1$.}

%% file: sec-security-proof.tex
\section{Security Model and Analysis}\label{sec-security}

In this paper, the described OMAP protocol for supporting the oblivious and fair exchange of privacy digital assets is based on extending data models and transaction definitions of Zcash protocol. It achieves fair and privacy-preserving exchange through orchestrated payment transactions. Since the OMAP algorithms are within the Zcash framework, most of the properties shown in Zcash and related publications still hold for OMAP ~\cite{miers2013zerocoin, kosba2015hawk,cryptoeprint:2016:061,cryptoeprint:2018:176}, for instance, completeness. The security proofs for Zcash are mostly applicable to OMAP with trivial extensions. Here the discussions focus on properties or requirements unique to the proposed exchange protocol. Note that it is assumed that the underlying distributed ledger system to support consensus and network communications is assumed to be secure and reliable, which is outside the scope of OMAP. We further assume that measures are taken to prevent attacks that may happen at different layers. Such attacks include but not limited to, for instance, 51\% attack, privacy compromise through analysis of network traffic patterns, attack to DNS, hard forks, quantum attack, etc. In addition, we assume that the underlying zero knowledge proving system is secure. 

\subsection{Requirements}

The original Zcash privacy coin defines security as: ledger indistinguishability, transaction non-malleability, and balance. 

\para{Ledger indistinguishability:} ledger L reveals no information to the adversary $\mathcal{A}$ beyond publicly disclosed information. 

\para{Transaction non-malleability:} no bounded adversary $\mathcal{A}$ can alter any of the data stored within a valid transaction. It prevents the adversary from modifying others' transactions before they are added to the ledger. 

\para{Balance:} no bounded adversary $\mathcal{A}$ can own more notes than what he minted or received via transactions from others.

Specific to OMAP, we define the following security requirements:  

\begin{itemize}

\item Fairness. When users agree to conduct an exchange of assets, represented as privacy notes that support multiple asset types, the operation leads to two possibilities: it fails that each one still has his/her note (asset), or it succeeds that each one possesses the other's note (asset type and amount). In other word, the exchange either completes or aborts. 

\item Privacy. Though all transaction information is stored on the decentralized ledger, other users should not be able to learn any useful information about the exchange transaction. For payment, Zcash is already shown to meet the requirement of transaction indistinguishability. OMAP extends JoinSplit transaction with multiple spending cases. Therefore, transaction indistinguishability is extended to cover the requirement that, no bounded adversary $\mathcal{A}$ can distinguish different scenarios of JoinSplit transactions. This means that OMAP ledger reveals no information to the adversary $\mathcal{A}$ of transactions so that $\mathcal{A}$ can tell if a JoinSplit transaction is a regular payment, exchange initialization, exchange cancellation, etc. 

\item Balance. No bounded adversary $\mathcal{A}$ can own more notes than what he minted, or received via direct payment from others or via exchange. 
\end{itemize}

\subsection{Analysis}

Next, we show that the constructed scheme satisfies all the features defined above, including, {\it fairness}, {\it indistinguishability}, and {\it balance}.


\para{Fairness:} This property requires that after an exchange is initiated, it either completes or aborts. At a high level, OMAP achieves fairness by reducing this requirement to assurance that double spending can be detected and prevented by the underlying ledger. According to OMAP, to start an exchange, Alice creates a pair of notes (one primary note and one sibling note). The primary note is shared between Alice and Bob, so either one of them can spend. If fairness is not satisfied, this means that the system reaches a state that either both Alice and Bob succeeded in spending the primary note; or none of them can spend the primary note. For the first case, it is clear that it is equivalent to double spending and is prohibited by the consensus mechanism of the underlying ledger. For the second case, if neither Alice nor Bob has spent the primary note yet, then one of them can spend in the future. In case, there is a time limit, Alice can eventually spend the primary note, which terminates the exchange operation. The likelihood that someone else besides Alice and Bob spends the primary note is negligible and ensured by the security of the base protocol. Therefore, one can conclude, that with probability 1-negl, the exchange will either succeed or terminate. 

\para{Privacy:}  In OMAP, the exchange is achieved through multiple JoinSplit transactions (shielded to shielded payment transactions). Although OMAP defines multiple JoinSplit transaction scenarios, there is only one combined circuit for proving ZK satisfiability of a JoinSplit transaction instance. This guarantees that different JoinSplit spending cases within an exchange operation or between exchange operations or between an exchange operation and regular payment should be indistinguishable. They are different ZK satisfying instances of the same proving circuit. If an adversary $\mathcal{A}$ can break the indistinguishability property, it means that $\mathcal{A}$ can distinguish different proofs, which contradicts to the assumption that the underlying zero-knowledge proving system is secure. 

\para{Balance:}  This requires that no bounded adversary $\mathcal{A}$ can own more notes than what he receives via direct payment or an exchange. Note that OMAP implements fair exchange using JoinSplit payment transactions. The original Zcash protocol already shows that the balance requirement is satisfied by JoinSplit transactions. OMAP introduces sibling note. Somehow, the asset in a sibling note can be viewed as borrowed from the system. A sibling note cannot be spent by itself.  According to JoinSplit algorithm, the condition to spend a sibling note is that either the associated primary note is spent already or it is spent together with the paired primary note in a JoinSplit transaction. This guarantees balance during the exchange operation. Assume that adversary $\mathcal{A}$  can spend a sibling note itself independent of the associated primary note. According to the OMAP JoinSplit ZK proof algorithm, this means that the adversary $\mathcal{A}$ must find another primary note with matching sequence number $S$. For a pair of primary note and sibling note, $S$ is created by a CRH using input value unique to the transaction. If adversary $\mathcal{A}$ can spend a sibling note by pairing it with another note, it means that adversary $\mathcal{A}$ can break the CRH by finding a collision, which contradicts the assumption that CRH is collision resistant.

%% file: sec-conclusion.tex
\section{Conclusion}\label{sec-conclusion}

Exchanging different types of crypto assets on distributed ledger/ blockchain is a natural extension of privacy-preserving cryptocurrency systems.
In response to this demand, we develop a novel multi-asset and privacy protected asset exchange framework on blockchain and apply zero-knowledge proof to enhance privacy protection for exchange operations.

In addition, we discuss the implementation of the unified zero-knowledge proving system used to support both payment transactions and multi-asset exchanges with privacy; and a formal definition of oblivious and privacy-preserving exchange scheme on the public ledger.

We extend the data formats and proving system of existing privacy coin to support the designed protocol. We have so far implemented circuits for handling extension and different payment spending scenarios in order to support fair asset exchanges using libsnark library. Furthermore, many new zero-knowledge proving systems are proposed recently, our current work includes experimenting and porting the scheme to zero-knowledge proving systems such as \textit{zk-STARK}~\cite{ben2018scalable} and \textit{Bulletproofs}~\cite{bunz2018bulletproofs}. 
Different from \textit{zk-SNARK}, recent schemes (e.g., \textit{Bulletproofs}) do not require a security setup. Implementing our multi-asset and oblivious exchange protocol over these new zero-knowledge proving systems may help increase both trustworthiness and performance of the proposed protocol.

%% file: privacy-swap.bbl
\begin{thebibliography}{10}

\bibitem{DBLP:journals/iacr/RonS12}
Dorit Ron and Adi Shamir.
\newblock Quantitative analysis of the full bitcoin transaction graph.
\newblock {\em {IACR} Cryptology ePrint Archive}, 2012:584, 2012.

\bibitem{van2013cryptonote}
Nicolas Van~Saberhagen.
\newblock Cryptonote v 2.0, 2013.

\bibitem{sasson2014zerocash}
Eli~Ben Sasson, Alessandro Chiesa, Christina Garman, Matthew Green, Ian Miers,
  Eran Tromer, and Madars Virza.
\newblock Zerocash: Decentralized anonymous payments from bitcoin.
\newblock In {\em 2014 IEEE Symposium on Security and Privacy}, pages 459--474.
  IEEE, 2014.

\bibitem{noether2015ring}
Shen Noether.
\newblock Ring signature confidential transactions for monero.
\newblock {\em IACR Cryptology ePrint Archive}, 2015:1098, 2015.

\bibitem{henry2018blockchain}
Ryan Henry, Amir Herzberg, and Aniket Kate.
\newblock Blockchain access privacy: challenges and directions.
\newblock {\em IEEE Security \& Privacy}, 16(4):38--45, 2018.

\bibitem{khalilov2018survey}
Merve Can~Kus Khalilov and Albert Levi.
\newblock A survey on anonymity and privacy in bitcoin-like digital cash
  systems.
\newblock {\em IEEE Communications Surveys \& Tutorials}, 20(3):2543--2585,
  2018.

\bibitem{pc-2}
Ethan Heilman, Leen AlShenibr, Foteini Baldimtsi, Alessandra Scafuro, and
  Sharon Goldberg.
\newblock Tumblebit: an untrusted bitcoin-compatible anonymous payment hub,
  2017.

\bibitem{pagnia1999impossibility}
Henning Pagnia.
\newblock On the impossibility of fair exchange without a trusted third party.
\newblock Technical report, 1999.

\bibitem{dziembowski2018fairswap}
Stefan Dziembowski, Lisa Eckey, and Sebastian Faust.
\newblock Fairswap: How to fairly exchange digital goods.
\newblock In {\em Proceedings of the 2018 ACM SIGSAC Conference on Computer and
  Communications Security}, pages 967--984. ACM, 2018.

\bibitem{bowe2018zexe}
S~Bowe, A~Chiesa, M~Green, I~Miers, P~Mishra, and H~Wu.
\newblock Zexe: Enabling decentralized private computation.
\newblock {\em IACR ePrint}, 962, 2018.

\bibitem{pc-13}
Neha Narula, Willy Vasquez, and Madars Virza.
\newblock zkledger: Privacy-preserving auditing for distributed ledgers.
\newblock In {\em 15th {USENIX} Symposium on Networked Systems Design and
  Implementation ({NSDI} 18)}, pages 65--80, Renton, WA, 2018. {USENIX}
  Association.

\bibitem{confidentail_assert}
Andrew Poelstra, Adam Back, Mark Friedenbach, Gregory Maxwell, and Pieter
  Wuille.
\newblock Confidential assets.
\newblock In Aviv Zohar, Ittay Eyal, Vanessa Teague, Jeremy Clark, Andrea
  Bracciali, Federico Pintore, and Massimiliano Sala, editors, {\em Financial
  Cryptography and Data Security}, pages 43--63, Berlin, Heidelberg, 2019.
  Springer Berlin Heidelberg.

\bibitem{miers2013zerocoin}
Ian Miers, Christina Garman, Matthew Green, and Aviel~D Rubin.
\newblock Zerocoin: Anonymous distributed e-cash from bitcoin.
\newblock In {\em Security and Privacy (SP), 2013 IEEE Symposium on}, pages
  397--411. IEEE, 2013.

\bibitem{tex2019dblp}
Rami Khalil, Arthur Gervais, and Guillaume Felley.
\newblock {TEX} - {A} securely scalable trustless exchange.
\newblock {\em {IACR} Cryptology ePrint Archive}, 2019:265, 2019.

\bibitem{das2019fastkitten}
Poulami Das, Lisa Eckey, Tommaso Frassetto, David Gens, Kristina
  Host{\'a}kov{\'a}, Patrick Jauernig, Sebastian Faust, and Ahmad-Reza Sadeghi.
\newblock Fastkitten: Practical smart contracts on bitcoin.
\newblock {\em IACR Cryptology ePrint Archive}, 2019:154, 2019.

\bibitem{goldwasser1985knowledge}
Shafi Goldwasser, Silvio Micali, and Charles Rackoff.
\newblock The knowledge complexity of interactive proof-systems.
\newblock In {\em Proceedings of the 7th annual ACM Symposium on Theory of
  Computing - STOC 1985}, pages 291--304. ACM, 1985.

\bibitem{goldreich1996construct}
Oded Goldreich and Ariel Kahan.
\newblock How to construct constant-round zero-knowledge proof systems for np.
\newblock {\em Journal of Cryptology}, 9(3):167--189, 1996.

\bibitem{blum1988non}
Manuel Blum, Paul Feldman, and Silvio Micali.
\newblock Non-interactive zero-knowledge and its applications.
\newblock In {\em Proceedings of the twentieth annual ACM symposium on Theory
  of computing}, pages 103--112. ACM, 1988.

\bibitem{groth2006perfect}
Jens Groth, Rafail Ostrovsky, and Amit Sahai.
\newblock Perfect non-interactive zero knowledge for np.
\newblock In {\em Annual International Conference on the Theory and
  Applications of Cryptographic Techniques}, pages 339--358. Springer, 2006.

\bibitem{de2001robust}
Alfredo De~Santis, Giovanni Di~Crescenzo, Rafail Ostrovsky, Giuseppe Persiano,
  and Amit Sahai.
\newblock Robust non-interactive zero knowledge.
\newblock In {\em Annual International Cryptology Conference}, pages 566--598.
  Springer, 2001.

\bibitem{gennaro2013quadratic}
Rosario Gennaro, Craig Gentry, Bryan Parno, and Mariana Raykova.
\newblock Quadratic span programs and succinct nizks without pcps.
\newblock In {\em Annual International Conference on the Theory and
  Applications of Cryptographic Techniques}, pages 626--645. Springer, 2013.

\bibitem{ben2013snarks}
Eli Ben-Sasson, Alessandro Chiesa, Daniel Genkin, Eran Tromer, and Madars
  Virza.
\newblock Snarks for c: Verifying program executions succinctly and in zero
  knowledge.
\newblock In {\em Advances in Cryptology--CRYPTO 2013}, pages 90--108.
  Springer, 2013.

\bibitem{ben2014succinct}
Eli Ben-Sasson, Alessandro Chiesa, Eran Tromer, and Madars Virza.
\newblock Succinct non-interactive zero knowledge for a von neumann
  architecture.
\newblock In {\em 23rd USENIX Security Symposium (USENIX Security 14)}, pages
  781--796, 2014.

\bibitem{hopwood2016zcash}
Daira Hopwood, Sean Bowe, Taylor Hornby, and Nathan Wilcox.
\newblock Zcash protocol specification.
\newblock {\em Tech. rep. 2016--1.10. Zerocoin Electric Coin Company, Tech.
  Rep.}, 2016.

\bibitem{cryptoeprint:2018:046}
Eli Ben-Sasson, Iddo Bentov, Yinon Horesh, and Michael Riabzev.
\newblock Scalable, transparent, and post-quantum secure computational
  integrity.
\newblock Cryptology ePrint Archive, Report 2018/046, 2018.
\newblock \url{https://eprint.iacr.org/2018/046}.

\bibitem{bunz2018bulletproofs}
Benedikt B{\"u}nz, Jonathan Bootle, Dan Boneh, Andrew Poelstra, Pieter Wuille,
  and Greg Maxwell.
\newblock Bulletproofs: Short proofs for confidential transactions and more.
\newblock In {\em 2018 IEEE Symposium on Security and Privacy (SP)}, pages
  315--334. IEEE, 2018.

\bibitem{cryptoeprint:2018:176}
Kamil Kluczniak and Man~Ho Au.
\newblock Fine-tuning decentralized anonymous payment systems based on
  arguments for arithmetic circuit satisfiability.
\newblock Cryptology ePrint Archive, Report 2018/176, 2018.
\newblock \url{https://eprint.iacr.org/2018/176}.

\bibitem{cryptoeprint:2017:602}
Sean Bowe, Ariel Gabizon, and Matthew~D. Green.
\newblock A multi-party protocol for constructing the public parameters of the
  pinocchio zk-snark.
\newblock Cryptology ePrint Archive, Report 2017/602, 2017.
\newblock \url{https://eprint.iacr.org/2017/602}.

\bibitem{cryptoeprint:2017:1050}
Sean Bowe, Ariel Gabizon, and Ian Miers.
\newblock Scalable multi-party computation for zk-snark parameters in the
  random beacon model.
\newblock Cryptology ePrint Archive, Report 2017/1050, 2017.
\newblock \url{https://eprint.iacr.org/2017/1050}.

\bibitem{kosba2015hawk}
Ahmed Kosba, Andrew Miller, Elaine Shi, Zikai Wen, and Charalampos Papamanthou.
\newblock Hawk: The blockchain model of cryptography and privacy-preserving
  smart contracts.
\newblock {\em University of Maryland and Cornell University}, 2015.

\bibitem{cryptoeprint:2016:061}
Christina Garman, Matthew Green, and Ian Miers.
\newblock Accountable privacy for decentralized anonymous payments.
\newblock Cryptology ePrint Archive, Report 2016/061, 2016.
\newblock \url{https://eprint.iacr.org/2016/061}.

\bibitem{ben2018scalable}
Eli Ben-Sasson, Iddo Bentov, Yinon Horesh, and Michael Riabzev.
\newblock Scalable, transparent, and post-quantum secure computational
  integrity.
\newblock {\em IACR Cryptology ePrint Archive}, 2018:46, 2018.

\end{thebibliography}
